\documentclass{ws-mpla}

\begin{document}

\markboth{E. Di Salvo, Z.J. Ajaltouni}
{Baryon Polarization, Phases of Amplitudes and Time Reversal Odd Observables}

\catchline{}{}{}{}{}

\title{BARYON POLARIZATION,\\
 PHASES OF AMPLITUDES \\
AND TIME REVERSAL ODD OBSERVABLES}

\author{\footnotesize E. DI SALVO\footnote{I.N.F.N. - Sez. Genova,
Via Dodecaneso, 33, 16146 Genova, Italy}}

\address{Laboratoire de Physique Corpusculaire de Clermont-Ferrand, \\
IN2P3/CNRS Universit\'e Blaise Pascal, F-63177 Aubi\`ere Cedex, France\\
disalvo@ge.infn.it}

\author{Z.J. AJALTOUNI}

\address{Laboratoire de Physique Corpusculaire de Clermont-Ferrand, \\
IN2P3/CNRS Universit\'e Blaise Pascal, F-63177 Aubi\`ere Cedex, France 
}

\maketitle

\pub{Received (Day Month Year)}{Revised (Day Month Year)}

\begin{abstract}
We elaborate two different methods for extracting from data the 
relative phase of two amplitudes of some sequential decays of 
baryons. We also relate this phase to a particular physical angle. 
Moreover we suggest how to infer some time reversal odd
observables - in particular, asymmetries - from data. Lastly, we 
comment on the standard model predictions of such asymmetries, 
showing that most of the decays considered are quite suitable 
for revealing possible new-physics contributions.

\keywords{Phase; rotation; time reversal odd.}
\end{abstract}

\ccode{PACS Nos.: 11.30.Er, 11.80.Et, 13.25.-k, 13.30.Eg}

\section{Introduction}
The search for new physics (NP) beyond the standard model (SM) is 
the main goal of present high energy physics. To this end, several 
facilities have been realized or planned. At the same time, different 
methods have been elaborated, in order to find hints to NP. In 
particular, as regards hadron decays, people try to single out and 
measure observables which are sensitive as much as possible to 
deviations from the SM. We mention among them the T-odd 
correlations\cite{va,wf,bl,bdl,bdl1} and asymmetries\cite{va,gr,aj,ajd,ajd1}, associated to kinematic variables which change sign at inverting all 
momenta and angular momenta involved\cite{va,wf}. These may be 
extracted, for example, from sequential weak two-body decays 
involving more spinning particles in the primary decay\cite{ajd1}. 
This kind of observable, wherever usable, looks more convenient than 
direct CP asymmetry, in that it does not depend so dramatically on 
the relative strong phase-shift of the interfering amplitudes\cite{va,bl,bdl}.
It has been employed in some experiments\cite{lhc,bel}. Also T-even
asymmetries may be quite helpful, although almost neglected in the
literature\cite{at,aj,daj,ajd1}.

We have studied in a previous paper\cite{ajd1} single and double 
asymmetries connected to weak sequential decays, in order to
infer suitable T-even and T-odd observables, sensitive to hints 
to NP. Here we focus on a particular kind of such decays, 
consisting of two successive weak processes of the type
\begin{equation}
J\to a ~~ b, ~~~~~~ a\to a_1 ~~ a_2. 
\label{sdc}
\end{equation}
Here $J,$ $a$ and $a_1$ are spin-1/2 baryons, $b$ is a
spin-0 or spin-1 meson and $a_2$ a spin-0 meson. We recall a 
previous result on these decays\cite{ajd1}, giving it a 
simple geometrical interpretation. Moreover we elaborate two 
different methods for extracting from data the relative 
phase of two decay amplitudes; to this end, we define a 
T-even and a T-odd single asymmetry. The convenience of
each method depends on the available statistics. We note that 
a nonzero T-odd asymmetry does not necessarily imply time 
reversal (TR) violation; however, if also data of the 
CP-conjugated decay may be analyzed, our methods allow to 
determine some observables which are odd under TR. Obviously, 
comparing the experimental results of these observables with 
their SM predictions, can in principle help detecting 
possible discrepancies, and therefore signatures of NP.
 
Here we list some decays of the type mentioned above, of current 
interest, to which our methods may be applied:
\begin{equation} 
\Lambda_b \to \Lambda (\Lambda_c^+) ~~~ P(V), \ ~~~~~~~~~ \
\Lambda_c^+ \to \Lambda ~~~ \pi^+, \ ~~~~~~~~~ \
\Xi  \to \Lambda ~~~ \pi. \label{lc}
\end{equation} 
Here $P(V)$ denotes a light, pseudoscalar (vector) meson.
We need also correlation with the secondary decay of the 
fermion, that is, {\it e. g.}, 
\begin{equation}
\Lambda \to p ~~~ \pi^-,
\end{equation}
or the second decay (\ref{lc}), if $\Lambda_c^+$ is observed in 
the primary decay. Data of such decays are available mainly
from Fermilab experiments; in particular, $\Lambda_b$ and 
$\Lambda_c$ decays have been investigated recently at 
CDF\cite{cdf,cdf4} and D0\cite{d0,abz}. For the hyperon 
decay we signal refs. \cite{hug,cha,bay}. The $\Lambda_b$ 
and $\Lambda_c$ decays have raised also a lively 
theoretical interest\cite{kld,ke1,arg,cgn}, as well 
as\cite{ke2}
\begin{equation}
\Sigma_b \to \Sigma_c ~~~ P(V), \ ~~~~~~~~~ \
\Omega_b \to \Omega_c  ~~~ P(V), \label{som}
\end{equation}
which may be detected at the new facilities, like LHCb.

The present note is organized as follows. In sect. 2 we 
introduce suitable reference frames for describing the 
sequential decay. Then we define operatively T-even and 
T-odd asymmetries, giving their analytical expressions 
in terms of physically interesting observables. Lastly, 
we expose several remarks, illustrating a geometrical 
interpretation of our results, and we describe our
first method for determining a T-even and a T-odd 
observable. In sect. 3 we suggest an alternative method, 
suitable in presence of a relatively scarce statistics. 
In sect. 4 we show how to determine real TR-odd 
observables, provided data of the CP-conjugated decay are 
available. Finally, sect. 5 is devoted to comments on 
the SM predictions of the asymmetries connected to 
the above mentioned decays, with a comparison with NP 
contributions, and to a short conclusion.

\section{T-odd and T-even Asymmetries} 

Here we introduce some T-odd and T-even asymmetries, which 
can be inferred from sequential decays of the type 
(\ref{sdc}). They are defined as
\begin{equation}
A = \frac{N(p>0)-N(p<0)}{N(p>0)+N(p<0)},
\label{pas}
\end{equation}
where $p$ is a given T-odd or T-even variable and 
$N(p>0)$ $[N(p <0)]$ the number of decays 
such that $p$ is positive (negative).

Our variables are conveniently defined in suitable 
reference frames, that we are going to define.
 
\subsection{Reference Frames}
First of all, we recall the definition of a canonical frame, at 
rest with respect to the parent resonance $J$:
\begin{equation}
{\hat{\bf y}} = \frac{{\bf p}_{in}}{|{\bf p}_{in}|},  ~~~~~~  
{\hat{\bf z}} =\frac{{\bf p}_{in}\times{\bf p}_J}
{|{\bf p}_{in}\times{\bf p}_J|},   ~~~~~~ 
{\hat{\bf x}} = {\hat{\bf y}}\times{\hat{\bf z}},
\label{reffr}
\end{equation}
where ${\bf p}_{in}$ and ${\bf p}_J$ are, respectively, the 
momenta of the initial beam  and of the resonance $J$ in the 
laboratory frame.

On the other hand, the secondary decay, {\it i.e.} of particle 
$a$, is more conveniently described in the helicity frame. To this
end we define the following three mutually orthogonal unit vectors:
\begin{equation}
{\hat{\bf e}}_L = \frac{{\bf p}_a}{|{\bf p}_a|},  ~~~~~~  
{\hat{\bf e}}_T = \frac{{\hat{\bf z}}\times{\hat{\bf e}}_L}
{|{\hat{\bf z}}\times{\hat{\bf e}}_L|},   ~~~~~~ 
{\hat{\bf e}}_N = {\hat{\bf e}}_T\times{\hat{\bf e}}_L,
\label{unvc}
\end{equation}
${\bf p}_a$ being the momentum of $a$ in the canonical 
frame.

\subsection{Asymmetries}

We define the T-odd and the T-even variable respectively as
\begin{equation}
p_N = {\bf p}_{a_1}\cdot{\hat{\bf e}}_N, 
\ ~~~~~~~ \
p_T = {\bf p}_{a_1}\cdot{\hat{\bf e}}_T, \label{toev}
\end{equation}
that is, the normal and transverse component of the momentum 
${\bf p}_{a_1}$ of particle $a_1$ in the rest frame of $a$. 
We denote the corresponding asymmetries, defined by Eq. 
(\ref{pas}), as $A_N$ and $A_T$. Their expressions 
read\cite{ajd1}\footnote{Eqs. (\ref{nasy}) and
(\ref{tasy}) differ from those in ref. \cite{ajd1}
by a factor of 2 at the denominator. This is due to 
a different definition of the polarization, coherent
with Eqs. (\ref{pdm}) below.}
\begin{eqnarray}
\Gamma(\Omega) A_N(\Omega) &=& \frac{1}{2\pi}
W \Delta a^{s_a} [\Re(\alpha_{1/2,0}
\alpha^*_{-1/2,0}){\cal P}_N + \Im(\alpha_{1/2,0}
\alpha^*_{-1/2,0}){\cal P}_T], \label{nasy}
\\
\Gamma(\Omega) A_T(\Omega) &=& \frac{1}{2\pi}
W \Delta a^{s_a} [\Im(\alpha_{1/2,0}
\alpha^*_{-1/2,0}){\cal P}_N - \Re(\alpha_{1/2,0}
\alpha^*_{-1/2,0}){\cal P}_T]. \label{tasy}
\end{eqnarray} 
Here $\Gamma(\Omega)$ is the differential decay width, $\Omega$ 
denoting the direction of the momentum of particle $a$ in the
canonical frame: as usual, we have set $\Omega \equiv (\theta,
\phi)$, where $\theta$ and $\phi$ are, respectively, the polar 
and azimuthal angle. Moreover
\begin{equation}
 W = |B(p^2_J)|^2|B(p^2_a)|^2|B(p^2_b)|^2
\end{equation}
and the $B(p^2)$'s are the relativistic Breit-Wigner functions 
of the resonances, normalized as
\begin{equation}
\int_0^{\infty}dp^2|B(p^2)|^2 = 1. \label{nrml}
\end{equation}
The $\alpha_{\lambda_a,\lambda_b}$'s are defined as
\begin{equation}
\alpha_{\lambda_a\lambda_b} = \frac{A^J_{\lambda_a\lambda_b}}
{\sqrt{\sum_{\lambda_a,\lambda_b}
|A^J_{\lambda_a\lambda_b}|^2}}, \label{nmam}
\end{equation}
the $A^J_{\lambda_a\lambda_b}$'s being the helicity decay 
amplitudes of the parent resonance and $\lambda_{a(b)}$
the helicity of particle $a (b)$. Similarly, we have set
\begin{equation}
\Delta a^{s_a} = \frac{1}{2}(a^{s_a}_+ - a^{s_a}_-) 
\ ~~~~~~~ \ \mbox{and} \ ~~~~~~~ \
a^{s_a}_{\pm} = \frac{|A^{s_a}_{\pm}|^2}{|A^{s_a}_+|^2+
|A^{s_a}_-|^2}, \label{sdch}
\end{equation}
where the $A^{s_a}_{\pm}$'s are the two helicity decay 
amplitudes of the baryon $a$. Furthermore,
\begin{equation}
\Gamma(\Omega) = \frac{1}{4\pi}W(1+{\cal P}_L\Delta G_L),
\end{equation}
where $\Delta G_L$ is defined as
\begin{equation} 
\Delta G_L = |\alpha_{1/2,0}|^2-|\alpha_{-1/2,0}|^2-
|\alpha_{1/2,1}|^2+|\alpha_{-1/2,-1}|^2,
\end{equation}
with $\alpha_{1/2,1}$ = $\alpha_{-1/2,-1}$ = 0
if the meson $b$ has spin 0. Lastly
\begin{equation}
{\cal P}_L = {\vec{\cal P}}\cdot{\hat{\bf e}}_L, \ ~~~~~~~ \ 
{\cal P}_N = {\vec{\cal P}}\cdot{\hat{\bf e}}_N, \ ~~~~~~~ \ 
{\cal P}_T = {\vec{\cal P}}\cdot{\hat{\bf e}}_T \label{plvc}
\end{equation}
are the components according to the helicity frame of the 
polarization vector ${\vec{\cal P}}$ of $J$. These 
depend on the angles $\theta$ and $\phi$, as we have
\begin{eqnarray}
{\hat{\bf e}}_L &\equiv& (sin\theta cos \phi, sin\theta sin \phi, 
cos \theta), \label{evl}
\\
{\hat{\bf e}}_T &\equiv& (-sin \phi, cos \phi, 0), ~~~~~~~ \ ~~~~~~~
\label{evt}
\\
{\hat{\bf e}}_N &\equiv& (-cos\theta cos \phi, -cos\theta sin \phi, 
sin\theta). \label{evn}
\end{eqnarray}

\subsection{Remarks}

~~~ A) Setting 
\begin{equation}
F = \alpha_{1/2,0}\alpha^*_{-1/2,0},
\end{equation}
Eqs. (\ref{nasy}) and (\ref{tasy}) imply that the quantities 
$\Re F$ and $\Im F$ are, respectively, T-even and T-odd,
since according to our definitions $A_T$ and ${\cal P}_T$ are 
T-even, while $A_N$ and ${\cal P}_N$ are T-odd. As explained 
in the introduction, a nonzero value of $\Im F$ would 
not necessarily imply TR violation (TRV), since it may be 
produced also by strong or electromagnetic final-state 
interactions (FSI), for example by  spin-orbit 
interaction\cite{br1,br2}, in the final state. 
Electromagnetic FSI are small and calculable\cite{ntrv}. 
On the contrary, strong FSI are sizable 
and difficult to calculate. We shall see below that, 
under some specific conditions and quite general 
assumptions, they may be disentangled experimentally 
from the effects of real TR.

B) Eqs. (\ref{nasy}) and (\ref{tasy}) can be rewritten as
\begin{equation}
A'_{\perp i}(\Omega) = C R_{ij}(-\Phi){\cal P} _{\perp j} 
(\Omega). \label{rot}
\end{equation}
Here we have introduced the two-dimensional vectors 
\begin{equation}
{\vec{\cal P}}_{\perp} \equiv ({\cal P}_N, {\cal P}_T)  
\ ~~~~~~~ \
\mbox{and} \ ~~~~~~~ \ {\vec{A'}}_{\perp} 
\equiv (A'_N, A'_T) \label{vct}
\end{equation}
in a plane perpendicular to ${\bf p}_a$, with
\begin{equation}
A'_N = A'_N(\Omega) = \Gamma(\Omega) A_N(\Omega),  
\ ~~~~~~~ \ 
A'_T = A'_T(\Omega) = -\Gamma(\Omega) A_T(\Omega).
\end{equation}

Moreover
\begin{equation}
\Phi = arg (F)
\end{equation}
and $R(-\Phi)$ is a rotation matrix such that the 
vector ${\vec{\cal P}}_{\perp}$ goes over into the 
direction of ${\vec A'}_{\perp}$; it is defined by 
$R_{NN}$ = $R_{TT}$ = $cos\Phi$, $R_{TN}$ = -$R_{NT}$ = 
$sin\Phi$. Lastly, $C$ is a constant,  
\begin{equation}
C = \frac{1}{2\pi} W \Delta a^{s_a} |F|.
\end{equation}
Therefore we conclude that the vectors (\ref{vct}) are 
connected to each other by a rotation by an angle opposite 
to the relative phase of the amplitude $A^J_{1/2,0}$ to 
$A^J_{-1/2,0}$.

C) These two-dimensional vectors are observable 
quantities, which can be inferred from experiment. 
In particular, thanks to Eqs. (\ref{plvc}) to 
(\ref{evn}), ${\cal P}_N$ and ${\cal P}_T$ are 
related to the components ${\cal P}_x, {\cal P}_y, 
{\cal P}_z$ of ${\vec{\cal P}}$ with respect 
to the canonical frame. These components can be 
deduced from the density matrix of the parent 
resonance $J$,
\begin{equation}
\rho_{\pm\pm} = \frac{1}{2}(1\pm {\cal P}_z),  
\ ~~~~~~~ \
\rho_{\pm\mp} = \frac{1}{2}({\cal P}_x \pm 
i{\cal P}_y), \label{pdm}
\end{equation}
which in turn can be determined from the differential 
width of the sequential decay. Therefore 
the modulus $|F|$ and the phase $\Phi$ - or 
equivalently $\Re F$ and $\Im F$ - can be deduced 
from Eqs. (\ref{nasy}) - (\ref{tasy}) or (\ref{rot}). 
$\Phi$ can be determined independently of the 
constant $C$: 
\begin{equation}
tan(\Phi) = \frac{{\vec{A'}}_{\perp}
\times{\vec{\cal P}}_{\perp}
\cdot {\hat{\bf e}}_L}{{\vec{A'}}_{\perp}
\cdot{\vec{\cal P}}_{\perp}}. \label{ang} 
\end{equation}
The ambiguity of $\Phi$ up to $\pi$, which derives 
from Eq. (\ref{ang}), can be resolved by 
introducing the solution into Eqs. (\ref{rot}).  

D) One can show that the asymmetries $A'_N$ and $A'_T$ 
do not coincide with the components of the polarization 
of the baryon $a$, when emitted in a given direction 
$\Omega$. However, the normal and transverse components 
of the polarization of the baryon $a$, averaged over
the whole solid angle, read
\begin{equation}
{\cal P}^a_N = \frac{\pi}{2}{\cal P}_z\Re F, 
\ ~~~~~~~~~~~~~ \
{\cal P}^a_T = -\frac{\pi}{2}{\cal P}_z\Im F, 
\label{pnt}
\end{equation}
as is straightforward to check. Note that 
${\cal P}^a_{N(T)}$ is T-odd (T-even), since 
${\cal P}_z = {\vec{\cal P}}\cdot {\hat{\bf z}}$ 
is T-odd according to the second Eq. (\ref{reffr}).
Result (\ref{pnt}) can be obtained also in a 
different way. Indeed, let us 
integrate the differential width\cite{ajd1} of 
the sequential decay (\ref{sdc}) over the solid 
angle $\Omega$ and over the polar angle of the 
momentum of $a_1$ in the helicity frame. We get 
\begin{equation}
\Gamma(\phi_1) = \frac{1}{2\pi}W[1+\frac{\pi^2}{8}
\Delta a^{s_a}{\cal P}_z\Re(F e^{i\phi_1})]. 
\label{dffi}
\end{equation}
Here $\phi_1$ is the azimuthal angle of the momentum 
of $a_1$ in the helicity frame. Comparing Eq.
(\ref{dffi}) with the general formula
\begin{equation}
\Gamma(\phi_1) = \frac{1}{2\pi}W[1+\frac{\pi}{2}
\Delta a^{s_a} \Re(\rho_{+-} e^{i\phi_1})],
\end{equation}
and taking account of Eqs. (\ref{pdm}),
leads again to formulae (\ref{pnt}). Note that, if the 
expression (\ref{dffi}) is used for fitting data, one 
can infer the values of ${\cal P}^a_N$ and 
${\cal P}^a_T$, and therefore $F$. 

Moreover Eqs. (\ref{pnt}) imply

- that $\Phi$ can be determined independently of 
${\cal P}_z$:
\begin{equation}
tan (\Phi) = -{\cal P}^a_T/{\cal P}^a_N; 
\end{equation}
   
- that Eqs. (\ref{nasy})-(\ref{tasy}) 
or (\ref{rot}) can be rewritten as
\begin{equation} 
\Gamma(\Omega) A_N(\Omega) = C'{\vec{\cal P}}^a_{\perp} 
\cdot  {\vec{\cal P}}_{\perp}, \ ~~~~~~~~~~~~~ \
\Gamma(\Omega) A_T(\Omega) = C'{\vec{\cal P}}_{\perp} 
\times  {\vec{\cal P}}^a_{\perp}\cdot {\hat{\bf e}}_L,
\end{equation}
where ${\vec{\cal P}_{\perp}}^a ~~~ \equiv ({\cal P}^a_N,
{\cal P}^a_T)$ and $C'$ = $2C/(\pi{\cal P}_z)$. 

E) Lastly, one has to observe that, in the production 
process of the parent resonance $J$, strong interactions 
contribute only to the $z$-component of the polarization 
${\vec{\cal P}}$, due to parity conservation. At sufficiently 
high energies, also weak interactions contribute, giving 
rise to nonzero values of the other components. 

\section{An Alternative Method}

Now we propose an alternative method, recommended when 
the available data is relatively poor. To this end, we 
define some integral quantities: 
\begin{eqnarray} 
B_{N(T)} &=& \int d\Omega A'_{N(T)}(\Omega), \label{int1}
\\
\Delta B^x_{N(T)} &=& \int_0^{\pi} sin\theta d\theta [\int_0^{\pi}-\int_{\pi}^{2\pi}]A'_{N(T)}(\Omega), 
\\
\Delta B^y_{N(T)} &=& \int_0^{\pi} sin\theta d\theta 
[\int_{-\pi/2}^{\pi/2}-\int_{\pi/2}^{3/2\pi}]A'_{N(T)}(\Omega).
\label{ints}
\end{eqnarray}
Experimentally, these observables amount to
\begin{eqnarray}
B_{N(T)} &=& \pm\frac{N(p_{N(T)}>0)- N(p_{N(T)}<0)}{N_{tot}},
\label{mom1}
\\
\Delta B^x_{N(T)} &=& \pm\frac{N(p_{N(T)} \cdot p^a_x >0)- 
N(p_{N(T)} \cdot p^a_x <0)}{N_{tot}}, \label{mom2}
\\
\Delta B^y_{N(T)} &=& \pm\frac{N(p_{N(T)} \cdot p^a_y >0)- 
N(p_{N(T)} \cdot p^a_y <0)}{N_{tot}}. \label{mom3}
\end{eqnarray}
Here $N_{tot}$ is the total number of events and the $+(-)$ 
sign refers to the index $N(T)$ of the above quantities. 
Moreover $ p^a_{x(y)}$ is the $x(y)$-component of ${\bf p}_a$ 
with respect to the canonical frame. Inserting Eqs. (\ref{rot}) 
into (\ref{int1})-(\ref{ints}) yields 
\begin{eqnarray}
B_N &=& \pi^2 C R_{NN}(-\Phi) {\cal P}_z, \ ~~~~~~~~ \ 
~~B_T = \pi^2 C R_{TN}(-\Phi) {\cal P}_z, \label{rel1}
\\
\Delta B^x_N &=& -8 C R_{NT}(-\Phi) {\cal P}_x, \ ~~~~~~~~ \
\Delta B^x_T = -8 C R_{TT}(-\Phi) {\cal P}_x, \label{rel2}
\\
\Delta B^y_N &=& +8 C R_{NT}(-\Phi) {\cal P}_y, \ ~~~~~~~~ \
\Delta B^y_T = +8 C R_{TT}(-\Phi) {\cal P}_y. \label{rel3}
\end{eqnarray}
We can determine the phase $\Phi$ and the modulus
$|F|$ from Eqs. (\ref{rel2})-(\ref{rel3}), since
the polarization of $J$ and the moments 
(\ref{mom1})-(\ref{mom3}) are experimental quantities.
In particular, the ratio
\begin{equation}
r = \frac{B_T}{B_N} = tan(\Phi) \label{tg2}
\end{equation}
allows to determine $\Phi$ independently of ${\cal P}_z$,
the ambiguity of the phase being resolved by the signs
of $B_N$ and $B_T$. Analogous ratios can be defined from 
Eqs. (\ref{rel2}) and (\ref{rel3}).
Therefore we can infer the relative phase of the 
amplitude $A^J_{1/2,0}$ to $A^J_{-1/2,0}$ by
just determining the moments (\ref{mom1})-(\ref{mom3}).

\section{Inferring Time Reversal Odd Observables}

As seen above, $\Re F$ is T-even, while $\Im F$ is T-odd. But 
\begin{equation}
\Re F = |F| cos \Phi, \ ~~~~~~~~~~~~~ \ \Im F = |F| sin \Phi. 
\label{trig}
\end{equation}
Therefore $\Phi$ itself must be T-odd. We set, without loss of 
generality,
\begin{equation}
\Phi = \Phi_e + \Phi_o, \label{evod}
\end{equation}
where $\Phi_{e(o)}$ is even (odd) under TR. Therefore, while 
$\Phi_o$ is sensitive to real TRV, $\Phi_e$ is a consequence 
of the fake\cite{bl,der,bi,si} T-odd effects caused by, say, 
spin-orbit interactions. Now we consider the decay which is 
CP-conjugated to (\ref{sdc}) and define 
\begin{equation}
{\bar F} = {\bar\alpha_{-1/2,0}} {\bar\alpha_{1/2,0}^*}
= |{\bar F}|e^{i{\bar\Phi}},
\end{equation}
where the barred symbols denote the quantities relative 
to this decay. In particular, the phase
\begin{equation}
{\bar\Phi} = {\bar\Phi_e} + {\bar\Phi_o}, \label{baroe}
\end{equation}
may be determined by a procedure analogous to 
the one suggested for the phase $\Phi$. But, if we 
assume the CPT symmetry, TR-even (TR-odd) terms are 
also CP-even (CP-odd). Then
\begin{equation}
{\bar\Phi_e} = \Phi_e, 
\ ~~~~~~~~~~~~~ \
{\bar\Phi_o} = -\Phi_o.\label{peod}
\end{equation} 
The phases $\Phi_e$ and $\Phi_o$ are observables, as they 
can be deduced from Eqs. (\ref{evod}), (\ref{baroe})
and (\ref{peod}). In this connection, it is useful to define
the TR-odd asymmetry   
\begin{equation}
A_{\Phi} = \frac{\Phi-{\bar\Phi}}{\Phi+{\bar\Phi}} = 
\frac{\Phi_o}{\Phi_e}. \label{ptd}    
\end{equation}
Note that, while $\Phi_o$ is TR-odd, $\Phi_e$ is a fake 
T-odd observable. Three more TR-odd asymmetries can be 
defined, {\it i. e.},
\begin{eqnarray}
A_F = \frac{|F|-|{\bar F}|}{|F|+|{\bar F}|}, \ ~~~~~~~~~ \
A_R = \frac{\Re{(F-{\bar F})}}{\Re{(F+{\bar F})}}, \ ~~~~~~~~~ \
A_I = \frac{\Im{(F-{\bar F})}}{\Im{(F+{\bar F})}}. \label{atr} 
\end{eqnarray}
Lastly we point out that the denominator of $A_I$, that is
\begin{equation} 
I_e =  \Im{(F+{\bar F})},
\end{equation}
is another fake T-odd observable. Of course, this kind of 
observables is as useful as real TR-odd ones\cite{dil,ddu} 
in the search for NP. 

\section{Comments on SM and NP Predictions of the TRV Asymmetries}

We devote this last section to comments on predictions of the 
TRV asymmetries defined above, for the decays (\ref{lc}) and 
(\ref{som}), both in the framework of the SM and 
according to NP contributions. 

First of all, we observe that the two asymmetries $A_N$ 
and $A_T$ are sensitive, not only to the relative phase $\Phi$ 
of the two reduced amplitudes
$\alpha_{1/2,0}$ and $\alpha_{-1/2,0}$, but also to the normal
and transverse component of the polarization of the parent
baryon, ${\cal P}_N$ and ${\cal P}_T$. As observed by other authors\cite{bl,bdl,bdl1,dl4}, this is an advantage over the 
$B$-decays to two vector mesons, which are insensitive 
to the spin of the $b$-quark, and whose triple product 
asymmetries undergo a strong suppression\cite{dl4}. 
Therefore, we can reasonably assume that the T-odd 
asymmetries defined in the present note are of the order 
of those calculated for the quarks involved in analogous 
decays\cite{bl,bdl1}.

Now let us examine in more detail the various decays 
considered in the introduction. Nonzero asymmetries     
are expected\cite{bl} in $\Lambda_b \to \Lambda 
(\pi^+\pi^-)_{\rho^0-\omega}$, where the two 
vector mesons interfere, and in $\Xi^0 \to \Lambda \pi^0$. 
Both decays derive contributions from the penguin diagram 
and from the color-suppressed tree diagram, which are 
kinematically different and have different weak 
phases\cite{dl4}, and therefore are suitable for 
producing nonzero TRV asymmetries of the type 
(\ref{ptd}) and (\ref{atr}). In particular, the tree 
diagram in the decay $\Lambda_b \to \Lambda 
(\pi^+\pi^-)_{\rho^0-\omega}$ is strongly Cabibbo 
suppressed; moreover the SM predictions of the TRV 
asymmetries related to this decay are of order\cite{arg} 
6 percent, while they are increased up to about
50 percent by NP contributions\cite{bdl}. On the contrary, 
the decay of $\Xi^0\to \Lambda \pi^0$ is characterized by 
a much more consistent contribution of the color-suppressed 
tree, only slightly Cabibbo suppressed; therefore it
may give rise to sizable TRV asymmetries. 

The other decays considered in the present note give rise 
to negligibly small TRV asymmetries according to the SM.
This is especially the case of those decays which involve a 
$b\to c$ or a $c\to s$ transition: indeed, they are driven 
essentially by a color-allowed and a color-suppressed tree 
diagram, which are kinematically different, but have 
approximately the same weak phase\cite{kld}. However, if we 
consider NP possible contributions to such decays, the 
predictions of the asymmetries are of order 10 to 20 
percent\cite{kld}.

To conclude, we have shown that some baryon sequential 
two-body decays can be exploited for extracting the relative 
phase $\Phi$ and the product of the moduli $F$ of the
reduced decay amplitudes $\alpha_{1/2,0}$ and 
$\alpha_{-1/2,0}$ of the parent baryon $J$. 
To this end, we have defined two asymmetries, which 
we regard as the components of the two-dimensional vector
${\vec{A'}}_{\perp}$, lying in a plane orthogonal to
the momentum of the intermediate baryon $a$. 
${\vec{A'}}_{\perp}$ is related 
to the projection of the polarization vector of $J$ onto
the same plane, by a rotation, whose angle is $-\Phi$. In 
order to determine $\Phi$ and $F$, two different 
methods are elaborated. If also data of 
the CP-conjugated decay may be analyzed, some important 
observables and TRV asymmetries can be inferred. They are 
especially sensitive to NP contributions in most of the 
decays considered, and therefore quite useful for 
singling out possible deviations from the SM.

\end{document}